\documentclass[preprint,aps]{revtex4}

\usepackage{graphicx}

\begin{document}

\title{Identification of Topological Surface State in PdTe$_2$ Superconductor by Angle-Resolved Photoemission Spectroscopy}

\author{Yan Liu$^{1,\sharp}$, Jianzhou Zhao$^{1,\sharp}$, Li Yu$^{1}$, Chengtian Lin$^{2}$, Aiji Liang$^{1}$, Cheng Hu$^{1}$, Ying Ding$^{1}$, Yu Xu $^{1}$, Shaolong He$^{1}$, Lin Zhao$^{1}$, Guodong Liu$^{1}$, Xiaoli Dong$^{1}$, Jun Zhang$^{1}$, Chuangtian Chen$^{3}$, Zuyan Xu$^{3}$, Hongming Weng$^{1}$, Xi Dai$^{1}$, Zhong Fang$^{1}$ and X. J. Zhou$^{1,4,*}$
}

\affiliation{
\\$^{1}$Beijing National Laboratory for Condensed Matter Physics, Institute of Physics, Chinese Academy of Sciences, Beijing 100190, China
\\$^{2}$Max-Planck-Institut f\"{u}r Festk\"{o}rperforschung, Heisenbergstrasse 1, 70569 Stuttgart, Germany
\\$^{3}$Technical Institute of Physics and Chemistry, Chinese Academy of Sciences, Beijing 100190, China
\\$^{4}$Collaborative Innovation Center of Quantum Matter, Beijing 100871, China
\\$^{\sharp}$These people contributed equally to the present work.
\\$^{*}$Corresponding authors: XJZhou@aphy.iphy.ac.cn
}
\date{May 25, 2015}

%Correspondence author. Email: XJZhou@aphy.iphy.ac.cn
% Email(The 1st author ): ly316803@163.com

%Tel: 13466705629 010-82648171

\begin{abstract}
High resolution angle-resolved photoemission measurements have been carried out on transition metal dichalcogenide PdTe$_2$ that is a superconductor with a T$_c$ at 1.7 K. Combined with theoretical calculations, we have discovered for the first time the existence of topologically nontrivial surface state with Dirac cone in PbTe$_2$ superconductor. It is located at the Brillouin zone center and possesses helical spin texture.  Distinct from the usual three-dimensional topological insulators where the Dirac cone of the surface state lies at the Fermi level, the Dirac point of the surface state in PdTe$_2$ lies deep below the Fermi level at $\sim$1.75 eV binding energy and is well separated from the bulk states. The identification of topological surface state in PdTe$_2$ superconductor deep below the Fermi level provides a unique system to explore for new phenomena and properties and opens a door for finding new topological materials in transition metal chalcogenides.\\


\end{abstract}

\maketitle

%%Introduction

Topological insulators are a new quantum state of matter with insulating bulk state and protected conducting edge or surface states that originate from its unique bulk band topology.\cite{RevHasan,RevZhang}  The concept of topological insulators has been extended  to topological superconductors that have attracted particular attention because they have a full pairing gap in the bulk and gapless surface states consisting of Majorana fermions.\cite{RevHasan,RevZhang,ZhangTSC}  The topological surface state exhibits unique electronic structure and spin texture which not only have potential applications in spintronics and quantum computing,\cite{RevQXL, RevMoore2} but also provide a promising platform for realizing exotic quantum phenomena,\cite{QiXL2008,RLi2010,QiXL2009,FuLPRL,YuR} including the revelation of Majorana fermions.\cite{MajoranaMajorana,WilczekMajorana}  The topological surface state has been observed mainly in the three-dimensional topological insulators Bi$_2$(Se,Te)$_3$ and related compounds, with a single Dirac cone observed near the Fermi level.\cite{YXia,HJZhang,YLChen} The discovery of topological surface state in new class of materials, particularly in superconducting materials,\cite{CavaCuBiSe,HasanCuBiSe} is of critical importance in realizing new quantum phenomena and potential applications.

The transition metal chalcogenides have been a rich playground to discover new materials with diverse physical phenomena and properties, such as the charge density wave,\cite{Moncton,Wilson1975} superconductivity,\cite{Smaalen,Wilson1969,Finlayson,Roberts,Sipos,Pyon,JJYang} extremely large magneto-resistance,\cite{Mazhar,PLCai} structural phase transition and strong spin-orbit interaction,\cite{Ootsuki2013,Ootsuki2014JPSJ,Ootsuki2014PRB} and materials like MoS$_2$ with application potentials.\cite{Novoselov,KFMak,Splendiani,Radisavljevic,Yoon,Duerloo,YiZhang}  However, there have been few reports related to the discovery of topological materials in the transition metal chalcogenides. In this Letter, we report the first time revelation of topological surface state in a transition metal dichalcogenide PdTe$_2$ that is also a superconductor with T$_c$$\sim$1.7 K.  By carrying out high resolution angle-resolved photoemission (ARPES) measurements on PdTe$_2$, combined with the theoretical calculations and topological invariant analysis, we have identified a topologically nontrivial surface state in PdTe$_2$ with a Dirac point deep below the Fermi level at a binding energy of $\sim$1.75 eV. The calculated spin texture of the surface state shows right-handed spin chirality for the upper Dirac branch and the left-handed for the lower one as in usual topological insulators. The discovery of topological surface state in PdTe$_2$ superconductor deep below the Fermi level provides a new platform to explore for new phenomena and properties related to topological materials.

%%Methods
High quality single crystals of PdTe$_2$ were obtained by the self-flux method. The ARPES measurements were performed on our lab photoemission system equipped with the Scienta R4000 electron energy analyzer.\cite{GDLiu,YLiu} We used the Helium discharge lamp with three kinds of photon energies: He I ($h\nu=21.218$ eV), He I $\alpha$($h\nu=23.087$ eV) and He II($h\nu=40.8$ eV), as the photon source for the ARPES measurements. The overall energy resolution used is 10 meV and the angular resolution is $\sim$0.3$^{\circ}$. The Fermi level (E$_F$) was referenced by measuring the Fermi edge of a clean polycrystalline gold electrically connected to the sample. The crystals were cleaved {\it in situ} and measured at a temperature of $T\sim 20$ K in vacuum with a base pressure better than 5$\times$10$^{-11}$ Torr.

The electronic structure of PdTe$_2$ was calculated by performing the first-principle calculation by WIEN2K package, which is based on the full potential linearized augmented plane wave (LAPW) method.\cite{Blaha} The lattice constants we used here are $a=b=4.036$ {\AA} and $c=5.13$ {\AA} for PdTe$_2$, taking a space group of $P\overline{3}m1$.\cite{McCarron,Finlayson,SJobic,WSKim}  The Brillouin zone integration was performed on a regular mesh of $14\times 14\times 9$ $k$ points. The muffin-tin radii ($R_{MT}$) of Pd and Te atoms are both 2.50 bohr. The largest plane-wave vector $K_{max}$ was given by $R_{MT}K_{max}=9.0$. The spin-orbit coupling (SOC) was included self-consistently in all calculations. To see the topological surface states, we construct a tight-binding model using the projected Wannier functions, which can reproduce the GGA band structure precisely.

%%Results and discussion

%%Figure 1
Figure 1 shows the crystal structure, the corresponding bulk and surface Brillouin zones, and the calculated and measured electronic structures of PdTe$_2$. The crystal structure of PdTe$_2$ (Fig. 1(a)) is polymeric CdI$_2$-type with the $P\bar{3}m1$ (No. 164) space group.\cite{McCarron,Soulard,SJobic,WSKim,JPJan}  The corresponding bulk Brillouin zone is a hexagonal prism, as shown  in Fig. 1b together with the projected (001) surface Brillouin zone. The measured Fermi surface of PdTe$_2$ (Fig. 1(c)) exhibits a complex topology with a three-fold symmetry\cite{YLiu}.  The overall band structure of PdTe$_2$ (Fig. 1(d)), measured along high symmetry directions and over a large energy range (0$\sim$6 eV), display multiple densely-arranged bands\cite{YLiu}. In order to understand the band structures we measured, we have performed the first-principle calculations with spin-orbit coupling  by GGA method\cite{Perdew}. Because of the strong three dimensionality of the electronic structures of PdTe$_2$,\cite{JPJan,GYGuo,Orders} the calculated band structure is presented in two different $k_z$ planes ${\it \Gamma} KM$ and $ALH$ (Fig. 1(e)). By comparing the measured band structure (Fig. 1(d)) with the calculations (Fig. 1e), we noticed the band structures seem more like that calculated in the $ALH$ plane, but still with a prominent discrepancy, as we have shown by overlying the theoretical calculated bands along $LAHL$ direction(black dashed lines in Fig. 1(d)) with the experiment data. The large discrepancy prompted us to check for possible reason on the inconsistency between the experiment and the calculations. Interestingly, we notice there is an obvious X-shaped band structure popping out at the Brillouin zone center(Fig. 1(d)) which is absent in the calculation in Fig. 1(e). We suspect whether there is a Dirac cone-like surface state here.

In the calculated band structure of PdTe$_2$ (Fig. 1(e)), there are eleven bands in the covered energy range -7 eV$\sim$3 eV that are all from strongly hybridized bands between Pd 4$d$- and Te 5$p$-orbitals\cite{JPJan,GYGuo,YLiu}. To study the topological nature of these bands, we use ''+" and ''-" to describe the parity for each band. There is a band inversion at $A$ point at $\sim$2 eV below the Fermi level E$_F$. We cut a curved Fermi level at 2 eV below $E_F$, and calculate the time-reversal invariant $Z_2$ parameter as shown in Table \ref{parity}. The results point to a possible existence of topological nontrivial surface state in PdTe$_2$. We then calculate the semi-infinite (001) surface state of a 40 layer slab based on the TB hamiltonian from Projected Wannier functions and presented in Fig. 1(f).  It exhibits a prominent surface state between the binding energy of 1.5 eV and 2.5 eV  at $\bar{\it \Gamma}$.  We notice that in the calculated energy window (3$\sim$-7 eV), all the surface states are merged together with the bulk states, except for this outstanding topological surface state near $\bar{\it \Gamma}$ at the binding energy of $\sim$1.75 eV, and some surface states across the Fermi level along the $\bar{\it \Gamma}-\bar{M}$ direction. Our measured ARPES data in Fig. 1(d) shows a remarkable agreement with the Dirac cone-like surface state at $\sim$2 eV binding energy from the surface state calculations (Fig. 1(f)). The identification of surface states in PdTe$_2$ solves the discrepancy found previously between the ARPES measurements and theoretical calculations.\cite{YLiu}

\begin{table}[!h]
\caption{The products of parity eigenvalues of the occupied states for the time reversal invariant momenta (TRIM) points, ${\it \Gamma}$, $M$, $A$, and $L$ in the Brillouin zone.}\label{parity}
\begin{tabular*}{0.35\textwidth}{@{\extracolsep{\fill}}ccccc}
  \hline\hline
  TRIM point & ${\it \Gamma}$ & $M$($\times$3) & $A$ & $L$($\times 3$) \\
  \hline
  Parity & $+$  & $-$ & $-$ & $-$\\
  \hline\hline
\end{tabular*}
\end{table}

%%Figure 2

To verify the surface state nature of the measured bands, we performed photon-energy dependent measurements to seek for its two-dimensional character: the surface state will not change with k$_z$.  We used three different photon energies to measure the same PdTe$_2$ sample: $h\nu$=21.218 eV, 23.087 eV and 40.8 eV. The measurements were carried out along two high-symmetry directions: $\bar{K}-\bar{\it \Gamma}-\bar{K}$ (upper panel in Fig. 2) and $\bar{M}-\bar{\it \Gamma}-\bar{M}$ (lower panel in Fig. 2).  Based on our previous work on the inner potential $V_0$=18.8 eV for PdTe$_2$,\cite{YLiu} the corresponding k$_z$ is estimated to be 5$\pi$/c for $h\nu$=21.218 eV, 5.13$\pi$/c for $h\nu$=23.087 eV and 6.22$\pi$/c for $h\nu$=40.8 eV.  This indicates that, for the He I (21.218 eV) and He I $\alpha$ (23.087 eV) lines, the ARPES detected planes are close to the $ALH$ plane, while for the He II (40.8 eV) line, it is close to the ${\it \Gamma}KM$ plane. It is known that the bulk electronic structure of PdTe$_2$ exhibits strong three-dimensionality,\cite{JPJan,GYGuo,Orders,YLiu} and small variation of k$_z$ can generate obvious change of the band structure\cite{YLiu}.  It is clear from Fig. 2 that the low energy features (0$\sim$1.5 eV binding  energy) and spectral intensity show strong variation with the photon energy change, reflecting their dominant bulk character. But the Dirac-cone-like bands near 1.75 eV in the Brillouin zone center appear to be quite robust against the change of the photon energy in terms of its energy position and intensity. This is consistent with its nature of being from the surface state. We also notice that the two high energy bands near 2.5 eV and 4.0 eV also do not show strong variation with photon energy change. As seen in Fig. 1(f), this can be understood because they have a large contribution from the surface states, especial for the band at $\sim$4 eV binding energy.

%%Figure 3
Figure 3 shows the measured constant energy contours at different binding energies for PdTe$_2$ (Fig. 3(a)), as well as the momentum dependence of the band structures (Fig. 3(b)), near the Dirac point at a binding energy of $\sim$1.75 eV in the Brillouin zone center. The constant energy contours (Fig. 3(a)) evolve strongly with the binding energy, from a circular ring at 1.5 eV, to a smallest spot near 1.8 eV, and gets larger again and becomes a large circular ring at 2.2 eV.  This is consistent with the typical behavior of a Dirac cone structure. Meanwhile, the momentum dependence of the band structure (Fig. 3(b)) also agrees with the Dirac cone picture. At the center point $\bar{\it \Gamma}$, it displayed two linear dispersion bands crossing at the Dirac point ($\sim$1.75 eV)(Fig. 3(b), left-most panel). When moving away from the $\bar{\it \Gamma}$ point, the bands gradually evolve into two parabolic-like bands without any intersection point. The upper band moves upward while the lower band shifts downward. All these findings provide strong evidence that this topological surface state possesses a typical Dirac-cone dispersion, as sketched in Fig. 3(c).

%%Figure 4
Figure 4 shows the topological surface state in PdTe$_2$ and its spin texture. Fig. 4(a) zooms in on the theoretically calculated surface state and Fig. 4(b) compares  the experimental data with the calculation. The experimentally observed Dirac point is at the binding energy of $\sim$1.75 eV and the agreement between the calculation and measurement is remarkable even in fine details. This surface state can be well-resolved because it stands out away from the bulk states, different from most of other surface bands that mix with the bulk bands.  As one important character of the topological surface state, we have calculated the spin texture for the upper and lower Dirac cone branches. The spin texture for the lower Dirac branch is exemplified by the spin texture at 0.3 eV below the Dirac point in Fig. 3(c).  The left-handed helical spin texture is expected for the lower Dirac cone branch.  The spin texture for the upper Dirac cone branch is right-handed, as shown by the spin texture at an energy of 0.06 eV above the Dirac point (Fig. 4(d)). Such spin texture is similar to that found in usual three-dimensional topological insulators, and can be tested in the future spin-resolved ARPES measurements.

%%Conclusion
In summary, by combining high resolution ARPES measurements with theoretical calculations, for the first time, we have established PdTe$_2$ superconductor as a new system that possesses topological surface state. The topological surface state in PdTe$_2$ is at the Brillouin zone center and is well separated from the bulk states. It possesses helical spin texture similar to that in three-dimensional topological insulators. Compared with the usual topological insulators where the Dirac cone lies near the Fermi level, our results provide a new case where the topological surface state is realized in a pristine superconductor instead of a topological insulator, and the Dirac cone of the surface state lies well below the Fermi level ($\sim$1.75 eV binding energy) instead of being at the Fermi level.  Further work needs to be done to explore whether such a unique topological surface state well below the Fermi level and in a superconductor can lead to new phenomena, properties and applications.  More work also needs to be done to investigate the nature of other surface states, particularly those crossing the Fermi level.  The identification of topological surface state in PdTe$_2$ also opens a door for searching for new topological materials  in transition metal chalcogenide systems.\\

\vspace{3mm}

\noindent {\bf Acknowledgement} XJZ thanks financial support from the NSFC (11190022), the MOST of China (973 program No: 2011CB921703 and 2011CBA00110), and  the Strategic Priority Research Program (B) of the Chinese Academy of Sciences (Grant No. XDB07020300).

\vspace{3mm}

\newpage

\begin{figure*}[tbp]
\begin{center}
\includegraphics[width=1.0\columnwidth,angle=0]{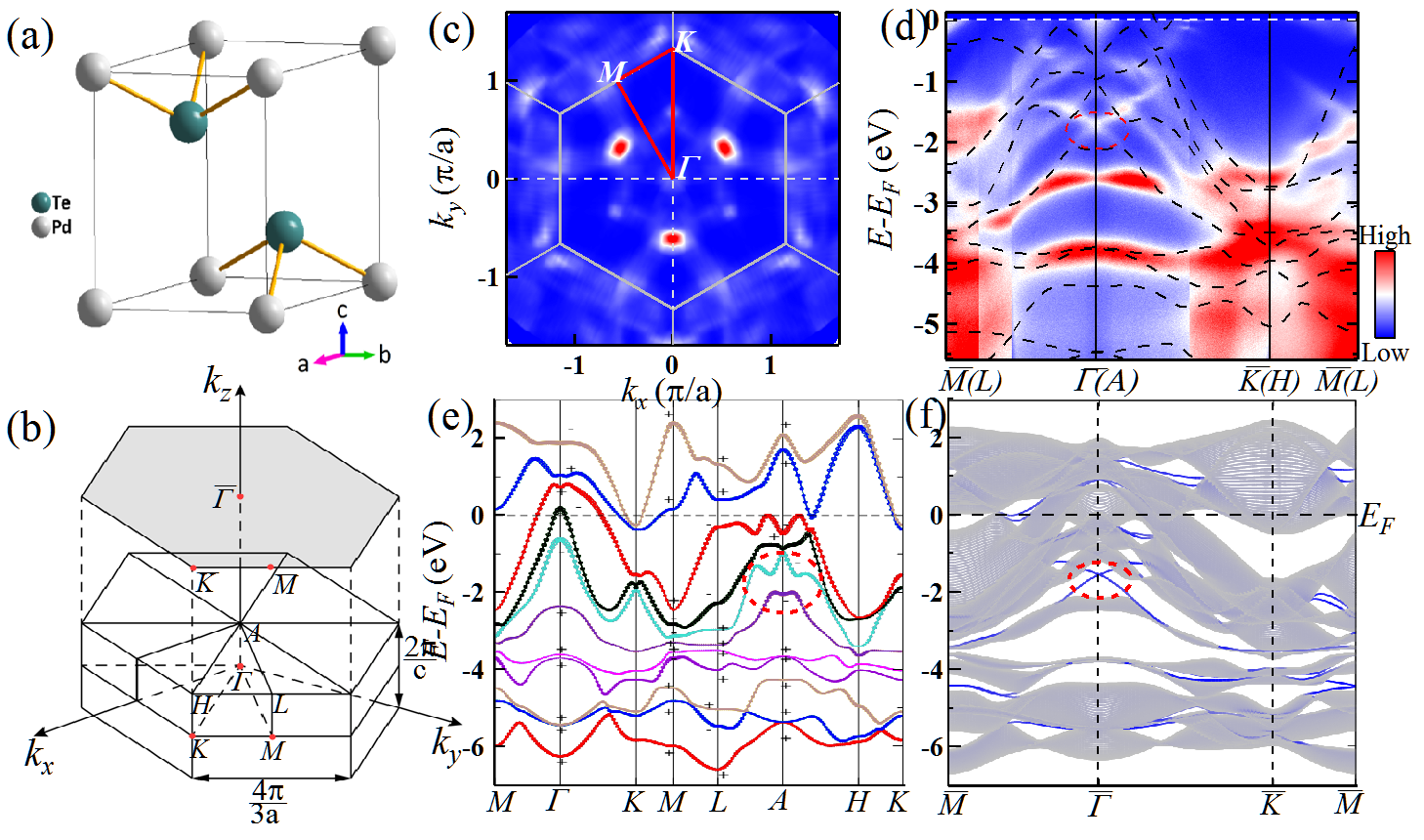}
\end{center}
\caption{Crystal structure, Brillouin zone, and experimentally measured and theoretically calculated electronic structures of PdTe$_2$. (a) Crystal structure (CdI$_2$-type) of PdTe$_2$ with $P$-$\bar{3}$$m$1 space group. (b) The bulk and (001) surface Brillouin zone for PdTe$_2$. (c) Measured Fermi surface of PdTe$_2$ by using Helium I line (photon energy: 21.218 eV ). (d) Band structure measured along high symmetry directions. The momentum cut locations are shown by red lines in (c). The black dashed lines on top of the photoemission image are the calculated band structures along $ALHL$ high-symmetry line shown in (e). The red dashed ellipse circles the Dirac cone-like topological surface state observed in experiment.  (e) The band structure with spin-orbit coupling by GGA calculation. The parities of each band at time reversal invariant momenta (TRIM) point are marked(``+'' for even, ``-'' for odd). The red dashed ellipse marks the band inversion at $A$ point at $\sim$-2 eV. (f) The surface state of PdTe$_2$ on the (001) surface. The surface states are obtained with a 40 layer slab on the basis of projected Wannier functions. The gray lines stands for the bulk state, while the blue lines are the surface state. The red dashed ellipse circles the calculated Dirac cone-like topological surface state.}
\end{figure*}

\begin{figure*}[tbp]
\begin{center}
\includegraphics[width=0.8\columnwidth,angle=0]{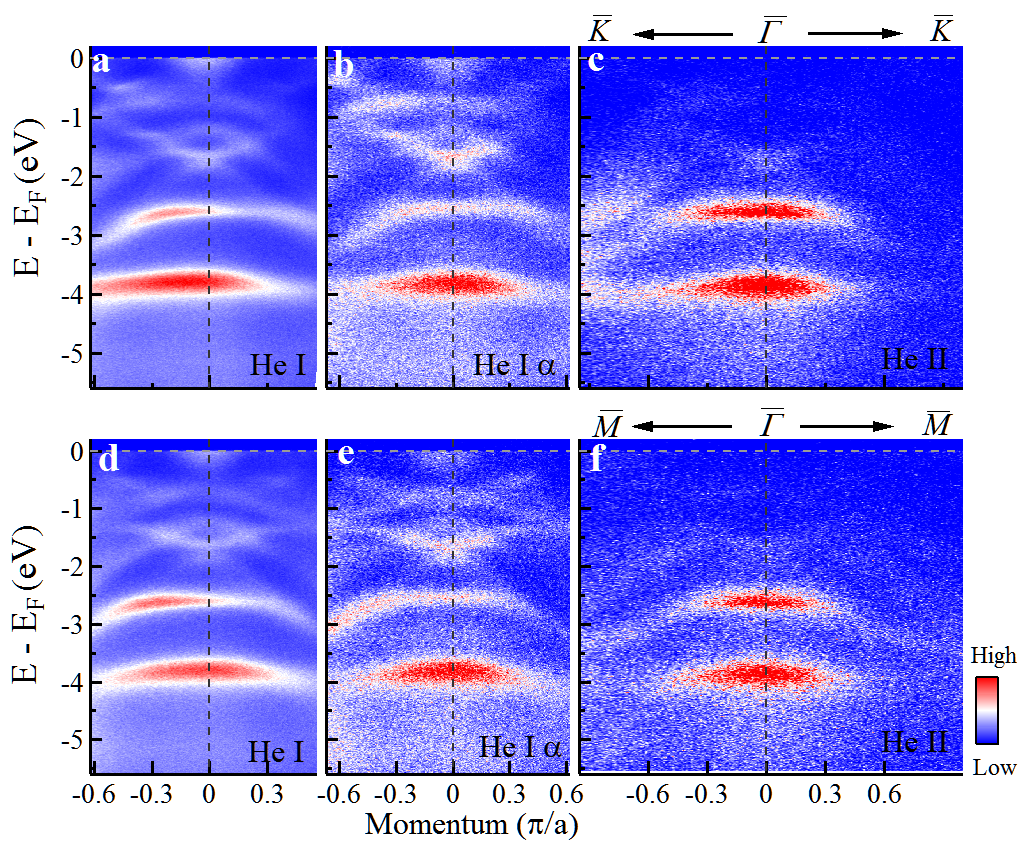}
\end{center}
\caption{Band structure measurements of PdTe$_2$ at $\sim$20 K by using three different photon energies. (a-c) are the band structures measured along $\overline{K}-{\it \overline{\Gamma}}-\overline{K}$ direction. (d-f) are the band structure measured along $\overline{M}-{\it \overline{\Gamma}}-\overline{M}$ direction. (a,d), (b,e) and (c,f) are the measurements by using Helium I (21.218 eV), Helium I$\alpha$ (23.087 eV) and Helium II (40.8 eV) as photon source, respectively. The low energy features between 0$\sim$1.5 eV binding energy show strong variation with the photon energy. While the Dirac-cone-like band near 1.75 eV at $\bar{\it \Gamma}$ stays robust. The two high energy bands near 2.5 eV and 4.0 eV binding energies show little variation either.}
\end{figure*}

\begin{figure*}[tbp]
\includegraphics[width=1.0\columnwidth,angle=0]{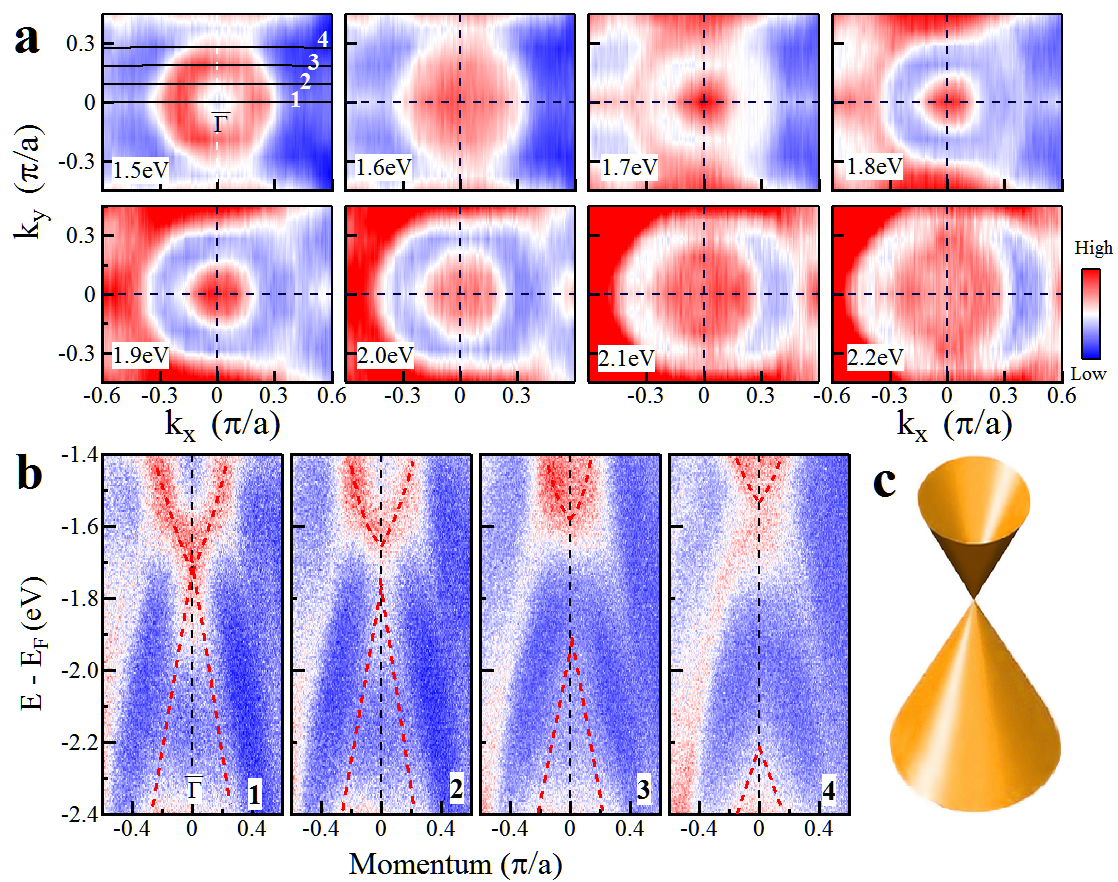}
\begin{center}
\caption{Constant energy contours and momentum dependence of the band structure of PdTe$_2$ around the Dirac point. (a) Constant energy contours at different binding energies from 1.5 eV (top-left panel) to 2.2 eV (bottom-right panel). The Dirac point energy is near the binding energy of 1.75 eV.  (b) Momentum dependence of the band structure along the four cuts show in (a)(black solid lines). The red dashed lines on top of the photoemission image represent guide to the eyes of the topological nontrivial surface state in PdTe$_2$. (c) A schematic Dirac Cone observed in PdTe$_2$.
}
\end{center}
\end{figure*}

\begin{figure*}[tbp]
\begin{center}
\includegraphics[width=0.8\columnwidth,angle=0]{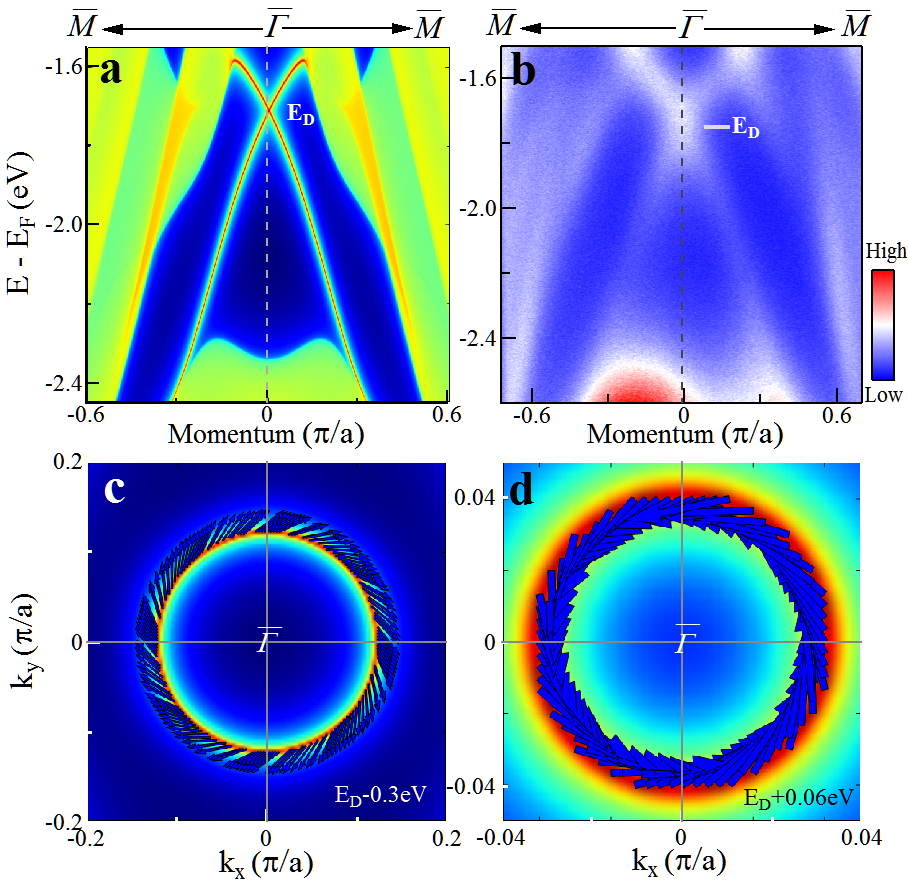}
\end{center}
\caption{Comparison between the calculated and measured topological surface state of PdTe$_2$ and the spin texture of the Dirac cone. (a) The zoom-in Dirac cone in the calculated surface state of PdTe$_2$ at ${\it \bar{\Gamma}}$ point. (b) The zoom-in measured Dirac cone surface state shown in a small binding energy window (1.5 eV-2.6 eV). (c) Calculated spin texture of lower Dirac cone branch at 0.3 eV below the Dirac point energy ($E_D$) shown in (a). (d) Calculated spin texture of the upper Dirac cone branch at 0.06 eV above the Dirac point energy $E_D$ shown in (a).
}
\end{figure*}


\begin {thebibliography} {99}


\bibitem{RevHasan} Hasan M Z and Kane C L 2010 \emph{Rev. Mod. Phys.} $\bf{82}$ 3045
\bibitem{RevZhang} Qi X L and Zhang S C 2011 \emph{Rev. Mod. Phys.} $\bf{83}$ 1057
\bibitem{ZhangTSC} Qi X L et al 2009 \emph{Phys. Rev. Lett.} $\bf{102}$ 187001
\bibitem{RevQXL} Qi X L and Zhang S C 2010 \emph{Phys. Today} $\bf{63}$ 33
\bibitem{RevMoore2} Moore J 2009 \emph{Nat. Phys.} $\bf{5}$ 378
\bibitem{QiXL2008} Qi X L, Hughes T L and Zhang S C 2008 \emph{Nat. Phys.} {\bf 4} 273
\bibitem{RLi2010} Li R, Wang J, Qi X L and Zhang S C 2010 \emph{Nat. Phys.} {\bf 6} 284
\bibitem{QiXL2009} Qi X L, Li R, Zang J and Zhang S C 2009 \emph{Science} {\bf 323} 1184
\bibitem{FuLPRL} Fu L and Kane C L 2008 \emph{Phys. Rev. Lett.} {\bf 100} 096407
\bibitem{YuR} Yu R et al 2010 \emph{Science} {\bf 329} 61
\bibitem{MajoranaMajorana} Majorana E 1937 \emph{Nuovo Cimento} $\bf{14}$ 171
\bibitem{WilczekMajorana} Wilczek F 2009 \emph{Nat. Phys.} $\bf{5}$ 614
\bibitem{YXia} Xia Y et al 2009 \emph{Nature Physics} {\bf 5} 398
\bibitem{HJZhang} Zhang H J et al 2009 \emph{Nature Physics} {\bf 5} 438
\bibitem{YLChen} Chen Y L et al 2009 \emph{Science} {\bf 325} 178
\bibitem{CavaCuBiSe} Hor Y S et al 2010 \emph{Phys. Rev. Lett.} {\bf 104} 057001
\bibitem{HasanCuBiSe} Wray L A et al 2010 \emph{Nat. Phys.} $\bf{6}$ 855
\bibitem{Moncton} Moncton D E, Axe J D and DiSalvo F J 1977 \emph{Phys. Rev. B} \textbf{16} 801
\bibitem{Wilson1975} Wilson J A, Salvo F J D and Mahajan S 1975 \emph{Adv. Phys.} \textbf{24} 117
\bibitem{Smaalen} Smaalen S V 2005 \emph{Acta Cryst.} \textbf{A61} 51
\bibitem{Wilson1969} Wilson J A and Yoffe A D 1969 \emph{Adv. Phys.} \textbf{18} 193
\bibitem{Finlayson} Finlayson T R 1986 \emph{Phy. Rev. B} \textbf{33} 2473
\bibitem{Roberts}  Roberts B W 1976 \emph{J. Phy. Chem. Ref. Data} \textbf{5} 581
\bibitem{Sipos} Sipos B et al 2008 \emph{Nature Materials} \textbf{7} 960
\bibitem{Pyon} Pyon S, Kudo K, and Nohara M 2012 \emph{J. Phys. Soc. Jpn.} \textbf{81} 053701
\bibitem{JJYang} Yang J J et al 2012 \emph{Phys. Rev. Lett.} \textbf{108} 116402
\bibitem{Mazhar} Ali M N et al 2014 \emph{Nature} \textbf{514} 205
\bibitem{PLCai} Cai P L et al 2014 arXiv:1412.8298 [cond-mat.mtrl-sci]
\bibitem{Ootsuki2013} Ootsuki D et al 2013 \emph{J. Phys. Soc. Jpn.} \textbf{82} 093704
\bibitem{Ootsuki2014JPSJ} Ootsuki K et al 2014 \emph{J. Phys. Soc. Jpn.} \textbf{83} 033704
\bibitem{Ootsuki2014PRB} Ootsuki D et al 2014 \emph{Phys. Rev. B} \textbf{89} 104506
\bibitem{Novoselov} Novoselov K S et al 2005 \emph{Proc. Natl. Acad. Sci. USA} \textbf{102} 10451
\bibitem{KFMak} Mak K F 2010 \emph{Phys. Rev. Lett} \textbf{105} 136805
\bibitem{Splendiani} Splendiani A 2010 \emph{Nano Lett.} \textbf{10} 1271
\bibitem{Radisavljevic} Radisavljevic B 2011 \emph{Nat. Nanotechnol.} \textbf{6} 147
\bibitem{Yoon} Yoon Y, Ganapathi K, and Salahuddin S 2011 \emph{Nano Lett.} \textbf{11} 3768
\bibitem{Duerloo} Duerloo KA N, Li Y and Reed E J 2014 \emph{Nature Communications} \textbf{5} 4214
\bibitem{YiZhang} Zhang Y et al 2014 \emph{Nature Nanotechnology} \textbf{9} 111
\bibitem{GDLiu} Liu G D et al 2008 \emph{Rev. Sci. Instrum.} \textbf{79} 023105
\bibitem{YLiu} Liu Y et al 2015 \emph{Chin. Phys. B} {\bf 24} 067401
\bibitem{Blaha} Blaha P et al 2001 \emph{WIEN2k, An Augmented Plane Wave + Local Orbitals Program for Calculating Crystal Properties} (Vienna: Vienna University of Technology).
\bibitem{McCarron} McCarron E, Korenstein R and Wold A 1976 \emph{Mater. Res. Bull.} \textbf{11} 1457
\bibitem{SJobic} Jobic S, Brec R and Rouxel J 1992 \emph{J. Solid State Chem.} \textbf{96} 169
\bibitem{WSKim} Kim W S, Chao G Y and Cabri L J 1990 \emph{J. Less-Common Met.} \textbf{162} 61
\bibitem{Soulard} Soulard C et al 2005 \emph{J. Solid State Chem.} \textbf{178} 2008
\bibitem{JPJan} Jan J P and Skriver H L 1977 \emph{J. Phys. F: Metal Phys.} \textbf{7} 1719
\bibitem{Perdew} Perdew J P,  Burke K and Ernzerhof M 1996 \emph{Phys.Rev.Lett.} {\bf 77} 3865
\bibitem{GYGuo} Guo G Y 1986 \emph{J. Phys. C: Solid State Phys.} \textbf{19} 5365
\bibitem{Orders} Orders P J et al 1982 \emph{J. Phys. F: Met. Phys.} \textbf{12} 2737



\end {thebibliography}
\end{document}